\documentclass[reprint,showpacs,amsmath,amssymb,aps,pra,superscriptaddress]{revtex4-2}
\usepackage[caption=false]{subfig}
\usepackage{graphicx}
\usepackage{dcolumn}
\usepackage{bm}
\usepackage{bbm}
\usepackage{amsthm}
\usepackage{mathtools}
\usepackage{physics}
\usepackage{binarytree}
\usepackage{tikz}
\usepackage{verbatim}
\usepackage[section]{placeins}
\usepackage[colorlinks, linkcolor=red, anchorcolor=blue, citecolor=green]{hyperref}
\usepackage{cleveref}
\crefname{equation}{Eq.}{Eqs.}
\Crefname{figure}{Fig.}{Figs.}

\newcommand{\sign}[1]{\mathrm{sgn}(#1)}
\begin{document}

\title{Disassociation of a one-dimensional cold molecule via quantum scattering}

\author{Wen-Liang Li}

\affiliation{Institute of Physics, Beijing National Laboratory for
  Condensed Matter Physics,\\Chinese Academy of Sciences, Beijing
  100190, China}

\affiliation{School of Physical Sciences, University of Chinese
  Academy of Sciences, Beijing 100049, China}

\author{Hai-Jing Song}

\affiliation{National Innovation Institute of Defense Technology, AMS, Beijing 100071, China}

\author{Tie-Ling Song}

\affiliation{Department of fundamental Science, Space Engineering University, Beijing 101416, China}

\author{D. L. Zhou} \email[]{zhoudl72@iphy.ac.cn}

\affiliation{Institute of Physics, Beijing National Laboratory for
  Condensed Matter Physics,\\Chinese Academy of Sciences, Beijing
  100190, China}

\affiliation{School of Physical Sciences, University of Chinese
  Academy of Sciences, Beijing 100049, China}

\date{\today}

\begin{abstract}
  Motivated by the recent experimental developments on ultracold molecules and atoms, we propose a simplest theoretical model to address the disassociation, reflection and transmission probability of a 1-dimensional cold molecule via quantum scattering. First, we give the Born approximation results in the weak interaction regime. Then, employing the Lippmann-Schwinger equation, we give the numerical solution and investigate the disassociation's dependence on the injection momentum and the interaction strengths. We find that the maximum disassociation rate has a limit as increasing the interaction strengths and injection momentum. We expect that our model can be realized in experiments in the near future.
\end{abstract}

\maketitle

\section{Introduction}
\label{sec:introduction}


Laser cooling makes atoms or molecules ultracold, e.g., the temperature may arrive at the regime of nano-Kelvin~\cite{lasercool,coolingmolecule,giorgini2008,blume2012,FITCH2021157},  which makes the emergence of quantum features of the atoms or molecules, which usually are hidden in the thermal noises from the environments. Thus the ultracold atoms or molecules becomes an ideal platform for investigation of fundamental quantum mechanics problems, quantum chemistry, precise quantum metrology, quantum simulations, and even quantum computing~\cite{safronova2018,pezze2018a,bloch2012,blackmore2018,gross2017,beterov2020,jaksch2000,bluvstein2022}.

Among these applications, ultracold chemistry is closely related with laser cooled atoms or molecules~\cite{heazlewood2021,liu2022}. Along this direction, one dimensional  ultracold atoms/molecules, which are formed by a tight confinement with a wave guide~\cite{PhysRevLett.81.938}, play a crucial rule due to its relatively simple theoretical model with rich physics~\cite{Guan_2022}.

Currently different kinds of molecules formed from several atoms have been investigated intensively in literature~\cite{yang2019,wang2021a,triatom,rydberg}. However, the converse process, i.e., the disassociation of molecules into atoms, deserves further studies to deepen its understandings. Here we propose a simplest theoretical model to address the dis-associative probability of a one-dimensional cold molecule, and investigate its dependence on the injection momentum and the interaction strengths, which can be arbitrarily tuned via the Feshbach resonance technique~\cite{kohler2006,chin2010}. Our results show that there is a limit of the maximum disassociation rate as increasing both the injection momentum and interaction strengths. 

This article is structured as follows: In \Cref{sec:hamitonian} we introduce our theoretical model of the scattering problem and give the Hamiltonian. In \Cref{sec:eigen-problem-h_0} we give the eigenstates and the in state of our scattering. Then we solve the model by applying Born approximation in \Cref{sec:born-appr-molecule} and integral equation method in \Cref{sec:integr-equat-meth}, and show our numerical results. Finally, we present our discussions and conclusions in \Cref{sec:disc-concl}. 

\section{The model}
\label{sec:hamitonian}

We consider a one-dimensional molecule, which is the unique weakly bound state formed by an attractive one-dimensional contact interaction. Then the one-dimensional molecule scatters with a heavy atom. The Hamiltonian of our system is modeled by
\begin{equation}
  \label{eq:1}
  H = \frac{p_{1}^{2}}{2 m_{1}} + \frac{p_{2}^{2}}{2 m_{2}} - \alpha
  \delta(x_{2}-x_{1}) + \gamma_{1} \delta(x_{1}) + \gamma_{2} \delta(x_{2}),
\end{equation}
where $\alpha, \gamma_{1}, \gamma_{2} > 0$. Here we assume that the position of the heavy atom is at zero, and the motion of the heavy atom is neglected.

To solve the scattering problem, we split the Hamiltonian into two parts:
\begin{equation}
  \label{eq:3}
  H = H_{0} + V
\end{equation}
where
\begin{align}
  H_{0} & = \frac{P^{2}}{2 M} + \frac{p^{2}}{2 \mu} - \alpha \delta(x),
  \\
  V & = \gamma_{1} \delta(X - r_{2} x) + \gamma_{2} \delta(X + r_{1} x),
\end{align}
with
\begin{align}
  \label{eq:2}
  r_{1} & = \frac{m_{1}}{M}, \\
  r_{2} & = \frac{m_{2}}{M}, \\
  M & = m_{1} + m_{2} = (r_1 + r_2)M, \\
  \mu & = \frac{m_{1} m_{2}}{M} = r_1 r_2 M, \\
  \kappa & = \sqrt{r_{1} r_2} = \sqrt{\frac{\mu}{M}}, 
\end{align}
\begin{align}  
  X & = \frac{m_{1} x_{1} + m_{2} x_{2}}{M} = r_1 x_1 + r_2 x_2, \\
  x & = x_{2} - x_{1}, \\
  P & = M \dot{X} = p_{1} + p_{2}, \\
  p & = \mu \dot{x} = r_{1} p_{2} - r_{2} p_{1}.
\end{align}

\section{The in state of our scattering}
\label{sec:eigen-problem-h_0}

In this section, we will examine the in state of our scattering. Let us start with the eigen problem of $H_{0}$, which can be divided into two parts:
\begin{align}
  \label{eq:18}
  H_{0} = H_{0}^{c} + H_{0}^{r},
\end{align}
where
\begin{align}
  \label{eq:19}
  H_{0}^{c} & = \frac{P^{2}}{2M}, \\
  H_{0}^{r} & = \frac{p^{2}}{2 \mu} - \frac{q}{\mu} \delta(x)
\end{align}
with $q=\mu\alpha$. Note that $H_0^c$ is the kinetic energy of the center of mass for the two atoms, and $H_0^r$ is the energy of their relative motion. Thus $[H_0^c, H_0^r]=0$, and the eigen problem of $H_0$ can be solved by finding the common eigenstates of $H_0^c$ and $H_0^r$.

The eigen equation of $H_0^c$ is given by
\begin{equation}
  \label{eq:42}
  H_0^c |P\rangle =\frac{P^{2}}{2M} |P\rangle,
\end{equation}
where the eigen wave function is
\begin{equation}
  \label{eq:20}
  \ip{X}{P} = \frac{1}{\sqrt{2\pi}} e^{iPX}.
\end{equation}

The eigen equation of $H_0^r$ is
\begin{align}
  \label{eq:43}
  H_0^r |\phi_b\rangle & = E_b |\phi_b\rangle, \\
  H_0^r |\phi_{p+}\rangle & = E_p |\phi_{p+}\rangle,
\end{align}
where $|\phi_b\rangle$ is the unique bound state with energy $E_b= - \frac{q^{2}}{2\mu}$, and the wave function for the bound state
\begin{equation}
  \label{eq:44}
  \langle x|\phi_{b}\rangle  = \sqrt{q} e^{-q |x|}.
\end{equation}
The eigenstate $|\phi_{p+}\rangle$ is the scattering state with respect to $H_0^r$ with energy $E_p=\frac{p^{2}}{2\mu}$, and the wave function is
\begin{equation}
  \label{eq:45}
  \langle x|\phi_{p+}\rangle =
  \begin{cases}
   &\frac{1}{\sqrt{2\pi}} \Bigl[\pqty{e^{i p x} + \frac{i q }{p - iq} e^{- i p x}} \theta(-x) \\
   &\quad + \pqty{\frac{p}{p - i q} e^{i p x}} \theta(x) \Bigr], \quad\quad p>0, \\
   &\frac{1}{\sqrt{2\pi}} \Bigl[\pqty{e^{i p x} + \frac{i q }{-p - iq} e^{- i p x}} \theta(x) \\
   &\quad + \pqty{\frac{-p}{-p - i q} e^{i p x}} \theta(-x) \Bigr], \quad p<0.
  \end{cases}
\end{equation}
Here we observe that $\langle x|\phi_{(-p)+}\rangle = \langle -x|\phi_{p+}\rangle$, i.e., $\langle -x|\phi_{p+}\rangle$ is also an eigenstate of $H_0^r$, which results from the symmetry of space inversion of $H_0^r$, i.e. the Hamiltonian is invariant under  $x\to -x$. In the Hilbert space of the relative motion, we can show the following complete relation
\begin{equation}
  \label{eq:7}
  \int_{-\infty}^{+\infty} dp \op{ \phi_{p+}} + \op{\phi_{b}} = 1.
\end{equation}

Now we are ready to  give the in state of our scattering
\begin{equation}
  \label{eq:8}
  |\Psi_{\text{in}}\rangle = |P\rangle\otimes|\phi_b\rangle \equiv |P, \phi_b\rangle,
\end{equation}
which describes a one-dimensional molecule in the bound state $|\phi_b\rangle$ scattering on the potential $V$ with the momentum of the mass center $P$.

\section{Born approximation in the molecule channel}
\label{sec:born-appr-molecule}

In this section, we will apply the Born approximation to our scattering problem. We start with the Lippmann-Schwinger equation:
\begin{align}
  \label{eq:4}
  \ket{\Phi_{P,b}^{+}}
  &  = \ket{P,\phi_{b}} + G^{+}(E_{P}) V
    \ket{P,\phi_{b}} \\
  & = \ket{P,\phi_{b}} + G_{0}^{+}(E_{P}) V
    \ket{\Phi_{P,b}^{+}},
\end{align}
where the Green function and the free Green function are given by
\begin{align}
  \label{eq:9}
  G^+(E) & =\frac{1}{E - H + i\epsilon}, \\
  G_0^+(E) & =\frac{1}{E - H_0 + i\epsilon}.
\end{align}
Therefore the $S$ matrix in the molecule channel is
\begin{align}
  \label{eq:5}
  \mel{Q,b}{S}{P,b}
  & = \ip{\Phi_{Q,b}^{-}}{\Phi_{P,b}^{+}} \\
  & = \delta(P-Q) -2i \pi \delta(E_{Q}-E_{P})
    \mel{Q,\phi_{b}}{V}{\Phi_{P,b}^{+}}.
\end{align}
The out scattering state in the molecule channel is
\begin{align}
  \label{eq:21}
  \ket{\Psi_{\text{out}}}_{b} & = \int_{-\infty}^{\infty} \dd Q \ket{Q,\phi_{b}}
                           \mel{Q,\phi_{b}}{S}{P,\phi_{b}} \nonumber\\
                         & = \pqty{1 - i \frac{2\pi M}{P}
                           \mel{P,\phi_{b}}{V}{\Phi_{P,b}^{+}}}
                           \ket{P, \phi_{b}} \notag\\
                           & \quad {} - i \frac{2\pi M}{P}
                           \mel{-P,\phi_{b}}{V}{\Phi_{P,b}^{+}} \ket{-P,
                           \phi_{b}}.
\end{align}
Then the reflection rate and the transmission rate for the molecule are
\begin{align}
  \label{eq:10}
  R_{b} & = \frac{4 \pi^{2} M^{2}}{P^{2}} \abs{\mel{-P,\phi_{b}}{V}{
          \phi^+_{P,b}}}^{2}, \\
  T_{b} & = 1 + \frac{4 \pi^{2} M^{2}}{P^{2}} \abs{\mel{P,\phi_{b}}{V}{
          \phi^+_{P,b}}}^{2} \notag \\ 
          &\quad + \frac{4\pi M}{P} \Im{\mel{P,\phi_{b}}{V
          }{\phi^+_{P,b}}}.
\end{align}
Therefore in the Born approximation up to second order of $V$:
\begin{align}
  \label{eq:25}
  R_{b} & = \frac{4 \pi^{2} M^{2}}{P^{2}} \abs{\mel{-P,\phi_{b}}{V}{P,
          \phi_{b}}}^{2}, \\
  T_{b} & = 1 + \frac{4 \pi^{2} M^{2}}{P^{2}} \abs{\mel{P,\phi_{b}}{V}{P,
          \phi_{b}}}^{2} \notag \\ 
          & \quad{} + \frac{4\pi M}{P} \Im{\mel{P,\phi_{b}}{V
          G_{0}^{+}(E_{P}) V}{P, \phi_{b}}}.\label{eq:22}
\end{align}
Note that
\begin{align}
  \label{eq:26}
  G_{0}^{+}(E_{p}) & = \frac{1}{E_{P} + E_{b} - H_{0} + i \epsilon}
                     \nonumber\\
                   & = \pv{\frac{1}{E_{P} + E_{b} - H_{0}}} - i \pi \delta(E_{P} +
                     E_{b} - H_{0}).
\end{align}
Thus
\begin{widetext}
\begin{align}
  \label{eq:27}
  & \Im{\mel{P,\phi_{b}}{V
    G_{0}^{+}(E_{P}) V}{P, \phi_{b}}} 
   = - \pi \mel{P,\phi_{b}}{V
    \delta(E_{P} + E_{b} - H_{0}) V}{P, \phi_{b}} \notag \\
  & = - \pi \int \dd{Q} \delta(E_{P} - E_{Q})
    \abs{\mel{Q,\phi_{b}}{V}{P,\phi_{b}}}^{2} - \pi \int \dd{Q} \int \dd{p} \delta(E_{P} + E_{b} - E_{Q} - E_{p})
    \abs{\mel{Q,\phi_{p+}}{V}{P,\phi_{b}}}^{2} \nonumber\\
  & = -\frac{\pi M}{P} \pqty{\abs{\mel{P,\phi_{b}}{V}{P,\phi_{b}}}^{2}
    + \abs{\mel{-P,\phi_{b}}{V}{P,\phi_{b}}}^{2}} \nonumber\\
  & \quad {} -  \int_{-p_{\text{max}}}^{p_{\text{max}}} \dd{p} \frac{\pi M}{Q(p)}
    \pqty{\abs{\mel{Q(p),\phi_{p+}}{V}{P,\phi_{b}}}^{2} +
    \abs{\mel{-Q(p),\phi_{p+}}{V}{P,\phi_{b}}}^{2} },
\end{align}
where \(Q(p)  = {\sqrt{p_{\text{max}}^2- p^{2}}}/{\kappa}\) with $p_{\text{max}}=\sqrt{\kappa^2 P^2 - q^2}$.

Hence, we find
\begin{equation}
  \label{eq:29}
  T_{b} = 1 - R_{b} - C_{nb},
\end{equation}
where
\begin{equation}
  \label{eq:30}
  C_{nb} = \frac{4 \pi^{2} M^{2}}{P} \int_{-p_{\text{max}}}^{p_{\text{max}}} \dd{p}
  \frac{\abs{\mel{Q(p),\phi_{p+}}{V}{P,\phi_{b}}}^{2} +
    \abs{\mel{-Q(p),\phi_{p+}}{V}{P,\phi_{b}}}^{2}}{Q(p)}.
\end{equation}
\end{widetext}
Eq.~\eqref{eq:29} implies that $C_{nb}$ is the disassociation rate, i.e., the rate that the molecule becomes two atoms after the scattering. In addition, only when $P>\frac{q}{\kappa}$ is $C_{nb}$ positive.

By detailed calculations, we obtain
\begin{equation}
  \label{eq:6}
  R_b = \frac{M^{2} q^{4}}{P^{2}} \pqty{
    \frac{\gamma_{1}}{q^2 +
          r_{2}^{2} P^{2}}  + \frac{\gamma_{2}}{q^2 +
          r_{1}^{2} P^{2} } }^{2},
\end{equation}
and
\begin{align}
  \label{eq:15}
  & \abs{\mel{Q,\phi_{p+}}{V}{P,\phi_{b}}}^{2}
   =  \pqty{\frac{q}{2\pi}}^{3} \frac{16{(P-Q)}^{2} p^{2}}{p^{2} +
   q^2} \nonumber\\
  & \quad\quad {} \left( \frac{1}{\bqty{p + (P-Q)r_{2}}^{2} + q^{2}} \pqty{    \frac{ r_{2}  \gamma_{1}}{
    \bqty{(P-Q)r_{2} - p}^{2} + q^2} }^{2}  \right. \nonumber\\
    & \quad\quad {} + \frac{1}{\bqty{p + (P-Q)r_{1}}^{2} + q^2} \pqty{    \frac{ r_{1}  \gamma_{2}}{
      \bqty{(P-Q)r_{1} - p}^{2} + q^2} }^{2}  \nonumber\\
  & \quad\quad {} + \frac{ r_{2}  \gamma_{1}}{
    \bqty{(P-Q)r_{2} - p}^{2} + q^2} \frac{ r_{1}
    \gamma_{2}}{ \bqty{(P-Q)r_{1} - p}^{2} + q^2}
    \nonumber\\
  & \quad\quad {} \left. \frac{2\bqty{p + (P-Q)r_{2}} \bqty{p +
    (P-Q)r_{1}} + 2 q^2}{\pqty{\bqty{p + (P-Q)r_{2}}^{2} + q^2} \pqty{\bqty{p + (P-Q)r_{1}}^{2} + q^2}} \right),
\end{align}
which can be inserted into Eq.~\eqref{eq:30} to numerically calculate the disassociation rate $C_{nb}$.

\begin{figure}[htbp]
  \centering
  \includegraphics[width=.48\textwidth]{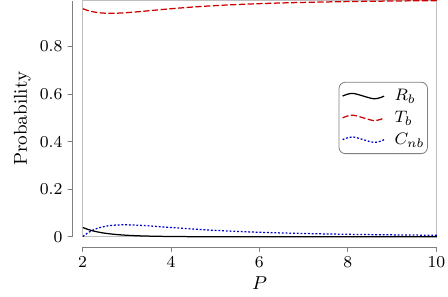}
  \caption{Disassociation under the Born approximation. Here the parameters are given by $m_1=m_2=1.0$, $\gamma_1=\gamma_2=0.2$, $\alpha=2.0$. }
  \label{fig:1}
\end{figure}

Now we are ready to present our numerical results on the transmission rate $T_b$, the reflection rate $R_b$, and the disassociation rate $C_{nb}$ in the first order Born approximation in Fig~\ref{fig:1}. In the case, the parameters are given by $m_1=m_2=1.0$, $\gamma_1=\gamma_2=0.2$, $\alpha=2.0$. Due to the energy conservation, only when the mass-center momentum $P>2$ does the disassociation process occur. With the increasing of the momentum $P$, the transmission rate $T_b$ increases while the reflection rate $R_b$ decreases. In particular, the disassociation rate $C_{nb}$ take its maximum $\simeq0.05$ at $P\simeq2.9$.

\section{Integral equation method}
\label{sec:integr-equat-meth}

Note that Born approximation is valid only when the momentum $P$ is large, and the interaction strengths $\gamma_1$ and $\gamma_2$ are small. To obtain more general information on the disassociation process, we may resort to the direct numerical solution of the Lippmann-Schwinger equation.

From Eqs.~\eqref{eq:25}\eqref{eq:22}, we need to calculate $V|\Phi^{+}_{P,b}\rangle$, which can be obtained from the Lippmann-Schwinger equation~\eqref{eq:4} and satisfies
\begin{equation}
  \label{eq:16}
  \left( 1 - V G_{0}^{+}(E_{P}) \right) V |\Phi^{+}_{P,b}\rangle = V |P, \phi_{b}\rangle.
\end{equation}
Therefore we arrives at the integral equation
\begin{equation}
  \label{eq:74}
  \begin{pmatrix}
    | \Phi^{1}\rangle \\ |\Phi^{2}\rangle
  \end{pmatrix}
  -
  \begin{pmatrix}
    G^{11} & G^{12}\\
    G^{21} & G^{22}
  \end{pmatrix}
  \begin{pmatrix}
    \gamma_{1} & 0\\
    0 & \gamma_{2}
  \end{pmatrix}
  \begin{pmatrix}
    | \Phi^{1}\rangle \\ |\Phi^{2}\rangle
  \end{pmatrix}
  =
  \begin{pmatrix}
    |\phi^{1}\rangle \\ |\phi_{2}\rangle
  \end{pmatrix},
\end{equation}
and the amplitudes of reflection rate and the transmission rate are given by
\begin{align}
  \label{eq:69}
    r_{b} & = i
            \frac{2\pi M}{P}
            \begin{pmatrix}
              \langle \psi^{1} | & \langle \psi^{2} |
            \end{pmatrix}
                      \begin{pmatrix}
                        \gamma_{1} & 0\\
                        0 & \gamma_{2}
                      \end{pmatrix}
                            \begin{pmatrix}
                              |\Phi^{1}\rangle \\ |\Phi^{2}\rangle
                            \end{pmatrix}, \\
  t_{b} & = 1- i
            \frac{2\pi M}{P}
            \begin{pmatrix}
              \langle \phi^{1}| & \langle \phi^{2}|
            \end{pmatrix}
                      \begin{pmatrix}
                        \gamma_{1} & 0\\
                        0 & \gamma_{2}
                      \end{pmatrix}
                            \begin{pmatrix}
                              |\Phi^{1}\rangle \\ |\Phi^{2}\rangle
                            \end{pmatrix},\label{eq:11}
\end{align}
where
\begin{align}
  \label{eq:73}
  \langle y|\Phi^{1}\rangle & = \langle r_{2} y,y|\Phi^{+}_{P,b}\rangle, \\
  \langle y| \Phi^{2}\rangle & = \langle- r_{1}y,y|\Phi^{+}_{P,b}\rangle, \\
  \langle y| \phi^{1}\rangle & = \langle r_{2}y,y|P,\phi_{b}\rangle, \\
  \langle y| \phi^{2}\rangle & = \langle - r_{1} y,y|P,\phi_{b}\rangle, \\
  \langle y|  \psi^{1} \rangle & = \langle r_{2}y,y|{-P},\phi_{b}\rangle, \\
  \langle y|  \psi^{2} \rangle & = \langle -r_{1}y,y|{-P},\phi_{b}\rangle, 
\end{align}
\begin{align}  
  \langle x|G^{11}|y\rangle & = \langle r_{2}x,x|G_{0}^{+}(E_{p})|r_{2}y,y\rangle, \\
  \langle x|G^{12}|y\rangle & = \langle r_{2}x,x|G_{0}^{+}(E_{p})|-r_{1}y,y\rangle, \\
  \langle x|G^{21}|y\rangle & = \langle- r_{1}x,x|G_{0}^{+}(E_{p})|r_{2}y,y\rangle, \\
  \langle x|G^{22}|y\rangle & = \langle -r_{1}x,x|G_{0}^{+}(E_{p})|-r_{1}y,y\rangle.
\end{align}

\subsection{Free Gree function}
\label{sec:free-gree-function}

To numerically evaluate the integral equation~\eqref{eq:74}, we need to calculate the free Green function
\begin{align}
  \langle X,x| G_{0}^{+}(E_{P})|Y,y\rangle  &= \langle X,x| \frac{1}{E_{P}+E_{b} - H_{0} + i \epsilon} |Y,y\rangle \notag \\
  & = G_{0}^{(I)} + G_0^{(II)}, \label{eq:76}
\end{align}
where
\begin{align}
  \label{eq:17}
    G_0^{(I)}               & =  \int_{-\infty}^{\infty} dQ \frac{\langle X,x|Q,\phi_{b}\rangle \langle Q, \phi_{b}|Y,y\rangle}{E_{P} - E_{Q} + i \epsilon}, \\
  G_0^{(II)} & = \int dQ \int dp \frac{\langle X,x|Q,\phi_{p^{+}}\rangle \langle Q, \phi_{p^{+}}|Y,y\rangle}{E_{P} - E_{Q} + E_{b} - E_{p} + i \epsilon}.
\end{align}
By detailed calculations, the free green function is given by
\begin{align}
  \label{eq:33}
  G_0^{(I)}   & = e^{-q(|x|+|y|)} \frac{-i M q e^{i  P|X-Y|}}{P}, \\
  G_0^{(II)} & = \frac{\kappa M}{2\pi i} \int_{-\infty}^{\infty} dp \Bigg[ \left(e^{ip|x-y|} + \frac{iq}{p-iq} e^{ip(|x|+|y|)} \right) \notag \\
  &\quad{} \times \frac{ e^{i \frac{|X-Y|}{\kappa} \sqrt{\kappa^{2} P^{2} - q^{2} - p^{2} }}}{\sqrt{\kappa^{2} P^{2} - q^{2} - p^{2} + i\epsilon}} \Bigg].
\end{align}
To further simplify the calculation of $G_0^{II}$, let 
\begin{align}
  \label{eq:65}
  p_0 & = \sqrt{|\kappa^2 P^2 - q^2|}, \\
  \sigma & = \sign{\kappa^2 P^2 - q^2}, \\
  q_0 & = \frac{q}{p_0}, \\
  \alpha & = p_{0} |x-y|, \\
  \beta & = p_{0} |X-Y|/\kappa, \\
  \eta & = p_0 (|x| + |y|), \\
  z & = \frac{p}{p_0}.
\end{align}
\begin{widetext}
Then the second term in the free Green function can be rewritten as
\begin{equation}
  \label{eq:66}
  G_0^{(II)}   =  \frac{\kappa M}{\pi i} \int_{0}^{\infty} dz \left(\cos(\alpha z) - \frac{q_0^2 \cos(\eta z) + q_0 z\sin(\eta z)}{z^2 + q_0^2} \right) \frac{ e^{i \beta \sqrt{\sigma  - z^{2} }}}{\sqrt{\sigma - z^{2}}}.
 \end{equation}
 It can be simplified as follows:

Case i: When $\kappa^{2}P^{2}-q^{2}<0$, $\sigma=-1$, and then
\begin{equation}
  \label{eq:67}
  G_0^{(II)}   =  - \frac{\kappa M}{\pi } \int_{0}^{\infty} du \left(\cos(\alpha \sinh(u)) - \frac{q_0^2 \cos(\eta \sinh(u)) + q_0 \sinh(u)\sin(\eta \sinh(u))}{\sinh(u)^2 + q_0^2} \right) e^{-\beta \cosh(u)}.
\end{equation}

Case ii: When $\kappa^{2}P^{2}-q^{2}>0$, $\sigma=1$, and then
\begin{align}
  \label{eq:75}
  G_0^{(II)}   & = - i\frac{\kappa M}{\pi } \int_{0}^{\frac{\pi}{2}} du \left(\cos(\alpha \sin{u}) - \frac{q_0^2 \cos(\eta \sin{u}) + q_0 \sin{u}\sin(\eta \sin{u})}{\sin{u}^2 + q_0^2} \right) \cos( \beta \cos{u})
       \nonumber\\
  & \phantom{=} + \frac{\kappa M}{\pi } \int_{0}^{\frac{\pi}{2}} du \left(\cos(\alpha \sin{u}) - \frac{q_0^2 \cos(\eta \sin{u}) + q_0 \sin{u}\sin(\eta \sin{u})}{\sin{u}^2 + q_0^2} \right) \sin( \beta \cos{u})
       \nonumber\\
  & \phantom{=} - \frac{\kappa M}{\pi } \int_{0}^{\infty} du \left(\cos(\alpha \cosh{u}) - \frac{q_0^2 \cos(\eta \cosh{u}) + q_0 \cosh{u}\sin(\eta \cosh{u})}{\cosh{u}^2 + q_0^2} \right) e^{- \beta \sinh{u}}.
\end{align}

Case iii: When $\kappa^2 P^2 -q^2=0$, $\sigma=0$, and then
\begin{equation}
  \label{eq:41}
  G_0^{(II)}    =  - \frac{\kappa M}{\pi } \int_{0}^{\infty} dp \left(\cos(p|x-y|) - \frac{q^2 \cos(p(|x|+|y|)) + q p\sin(p(|x|+|y|))}{p^2 + q^2} \right) \frac{ e^{- \frac{p|X-Y|}{\kappa}}}{p}.
\end{equation}
\end{widetext}

\subsection{Numerical results}
\label{sec:numerical-results}

\begin{figure}[htbp]
  \centering
  \includegraphics[width=.48\textwidth]{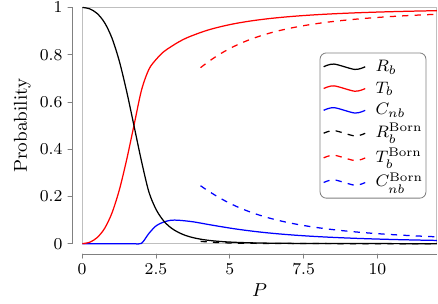}
  \caption{Disassociation rate from numerical solution of the integral equation compared with Born approximation, where the parameters are given by $m_1=m_2=1.0$, $\gamma_1=\gamma_2=0.5$, $\alpha=2.0$.}
  \label{fig:2}
\end{figure}

Now we are ready to perform the numerical solution of the integral equation~\eqref{eq:74} to obtain $|\Phi^1\rangle$ and $|\Phi^2\rangle$, and calculate the reflection rate $R_b$ and the reflection rate $T_b$ via Eqs.~\eqref{eq:69}\eqref{eq:11}. Then the disassociation rate can be obtained by $C_{nb}=1-R_b-T_b$ in Fig~\ref{fig:2}, where the parameters are given by $m_1=m_2=1.0$, $\gamma_1=\gamma_2=0.5$, $\alpha=2.0$. Compared with the case calculated in the Born approximation, we take larger scattering strengths $\gamma_1$ and $\gamma_2$ while keeping the other parameters invariant. As expected, the disassociation channel opens only when the mass-center momentum $P>2$.  With the increasing of the momentum $P$, the transmission rate $T_b$ increases while the reflection rate $R_b$ decreases. The disassociation rate $C_{nb}$ takes its maximum $\simeq0.1$ at $P\simeq3.2$. We also show the Born approximation results in the same parameter setting, which become increasingly accurate with the integral results as \(P\) increasing, just as one can expect. 

\begin{figure*}[htbp]
  \vspace{-2em}
  \centering
  \subfloat[]{
    \centering
    \label{fig:3a}
    \includegraphics[width=.48\textwidth]{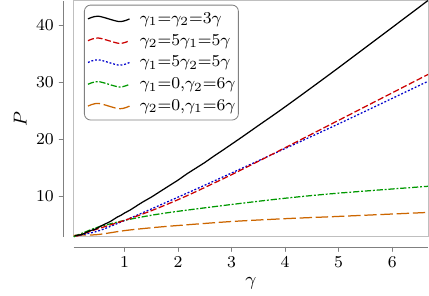}
    }
  \hfil
  \subfloat[]{
    \centering
    \label{fig:3b}
    \includegraphics[width=.48\textwidth]{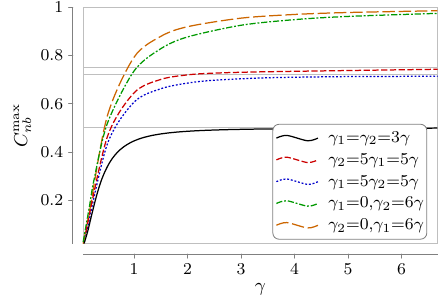}
    }
    \caption[]{ The parameters of the solid black lines are $m_1=m_2=1.0$, $\gamma_1=\gamma_2=\gamma$, $\alpha=2.0$. And the parameters of dashing lines are \(m_1=0.5, m_2=1.5, \alpha = 2.0\). \subref{fig:3a} The conditions of center-of-mass momentum \(P\) and the interaction strengths \(\gamma\) when the disassociation rate \(C_{nb}\) takes its maximum \(C_{nb}^{\mathrm{max}}\). \subref{fig:3b} The maximum of disassociation rate \(C_{nb}^{\mathrm{max}}\) changing with the interaction strengths \(\gamma\).  \label{fig:3}}
\end{figure*}

\begin{figure*}[htb]
  \vspace{-2em}
  \centering
  \subfloat[]{
    \centering
    \label{fig:4a}
    \includegraphics[width=.48\textwidth]{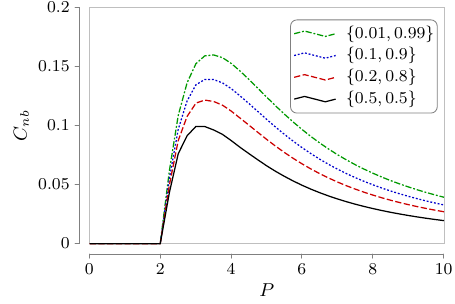}
  }
  \hfil
  \subfloat[]{
    \centering
    \label{fig:4b}
    \includegraphics[width=.48\textwidth]{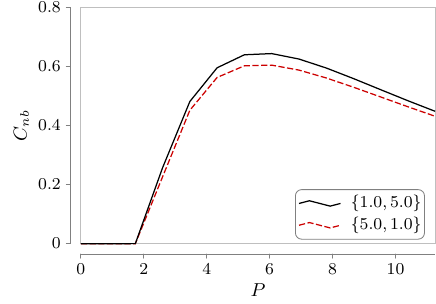}
  }
  \caption[]{The disassociation rate with different interaction strengths \(\Bqty{\gamma_1,\gamma_2}\), where the parameters are given by \subref{fig:4a} $m_1=m_2=1.0$, $\alpha=2.0$ and \subref{fig:4b} $m_1=0.5, m_2=1.5$, $\alpha=2.0$. \label{fig:4}}
\end{figure*}

We also care about that how the parameters influence the maximum of the disassociation rate. The disassociation rate depends on the mass of each particle, the interaction strengths \(\Bqty{\gamma_1,\gamma_2}\) and center-of-mass momentum \(P\) for a fixed bound strength \(\alpha\). In \Cref{fig:3}, we show when the disassociation rate takes its maximum \(C_{nb}^{\mathrm{max}}\) under different parameter settings. The solid black lines in \Cref{fig:3} show \(C_{nb}^{\mathrm{max}}\) with equal interaction strengths \(\gamma_1=\gamma_2=\gamma\), and equal mass \(m_1 = m_2 = 1.0\), while the dashing lines show \(C_{nb}^{\mathrm{max}}\) with \(m_1 = 0.5, m_2 = 1.5\) and different interaction strengths. The bound strength is \(\alpha = 2.0\). \Cref{fig:3a} shows the conditions of \(P\) and \(\gamma\) when \(C_{nb} = C_{nb}^{\mathrm{max}}\), which means that in order to reach the maximum disassociation rate, one should increase both \(P\) and \(\gamma\) following the relations revealed in \Cref{fig:3a}. \Cref{fig:3b} gives the values of \(C_{nb}^{\mathrm{max}}\) under different parameter settings changing with the interaction strength \(\gamma\), from which we can see that they increase as \(\gamma\) increasing and asymptotically reach some limits. For equal mass and equal interaction strengths, the limit of \(C_{nb}^{\mathrm{max}}\) is 0.5. For \(\gamma_1= 5 \gamma_2\), the limit is about 0.72, and for \(\gamma_2 = 5 \gamma_1\), the limit is about 0.75. For \(\gamma_1=0\) or \(\gamma_2 =0\), the limit approximates to 1. In conclusion, if one want to reach higher disassociation rate, one would tune stronger interaction strengths and center-of-mass momentum following some similar relations given in \Cref{fig:3a} and a larger difference between interaction strengths \(\gamma_1\) and \(\gamma_2\). In fact this maximum value \(C_{nb}^{\mathrm{max}}\) is irrelevant to the coupling strength \(\alpha\) in this situation because this can be reduced to a scaling problem. 

While for different interaction strengths (\(\gamma_1 \neq \gamma_2\)), one would suppose that larger difference between \(\gamma_1\) and \(\gamma_2\) induce larger disassociation rate. \Cref{fig:4} shows more details of the effect, where we keep \(\gamma_1 + \gamma_2 = 1.0\) in \Cref{fig:4a} to see the main influence of the difference between \(\gamma_1\) and \(\gamma_2\). \Cref{fig:3b,fig:4b} also show that lighter particle in the molecule with weaker interaction strength has higher disassociation rate than that of lighter particle with stronger interaction strength. 

\begin{figure}[htb]
  \centering
  \includegraphics[width=.48\textwidth]{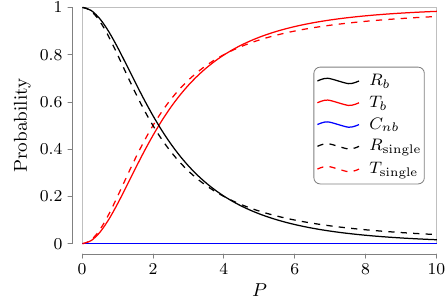}
  \caption{Reflection and transmission rates of the molecule compared with a single particle, where the parameters are give by $m_1=m_2=1.0$, $\gamma_1=\gamma_2=0.5$, $\alpha=12.0$. \label{fig:single}}
  \vspace{-1em}
\end{figure}

When the coupling of the molecule is strong enough, in the regime of low injection center-of-mass momentum \(P\) the molecule would not disassociate and behave as a single particle. We know the reflection rate \(R_{\mathrm{single}}\) and transmission rate \(T_{\mathrm{single}}\) of a single particle scattered by a \(\delta\) potential, which is a kind of quantum tunneling~\cite{griffiths2018}, and in our problem: 
\begin{align}
  R_{\mathrm{single}} & = \frac{M^2(\gamma_1 + \gamma_2)^2}{P^2 + M^2(\gamma_1 + \gamma_2)^2}, \label{eq:R_single}\\
  T_{\mathrm{single}} & = \frac{P^2}{P^2 + M^2(\gamma_1 + \gamma_2)^2}. \label{eq:T_single}
\end{align}
\Cref{fig:single} shows the reflection and transmission rates of the molecule compared with a single particle for \(P<10\), where the parameters are give by $m_1=m_2=1.0$, $\gamma_1=\gamma_2=0.5$, $\alpha=12.0$. 

\section{Discussion and conclusion}
\label{sec:disc-concl}

In this paper, a simple model with contact interactions, which contains the basic process of disassociation of a  one-dimensional molecule, is proposed to describe the corresponding system of ultracold atoms. The first order Born approximation is made to obtain the basic physical picture of the process: only when the kinetic energy associated with the injection center-of-mass momentum $P$ is larger than the ionization energy can the disassociation process occur. To further validate this picture, we develop the numerical method to solve the integral equation of quantum scattering. With the increases of the interaction strengths and the injection center-of-mass momentum, the maximum disassociation rate will increase, With larger difference of the interaction strengths the disassociation rate will increase. And under different parameter settings, the maximum disassociation rate has different limits as increasing the interaction strengths and injection momentum. We expect that our model can be realized in the experiment of ultracold atoms and molecules in the near future.

\begin{acknowledgments}
  This work is supported by National Key Research and Development Program of China (Grants No. 2021YFA0718302 and No. 2021YFA1402104), National Natural Science Foundation of China (Grant No. 12075310), and the Strategic Priority Research Program of Chinese Academy of Sciences (Grant No. XDB28000000).
\end{acknowledgments}

\bibliographystyle{apsrev4-2}
\bibliography{dimerref.bib}

\end{document}